\documentclass[conference]{IEEEtran}
\IEEEoverridecommandlockouts

\usepackage{amsmath,amssymb,amsfonts}
\usepackage{booktabs}
\usepackage{algorithmic}
\usepackage{graphicx}
\usepackage{textcomp}
\usepackage[table,xcdraw]{xcolor}
\usepackage{hyperref}
\hypersetup{%
  colorlinks=true,
  linkcolor=red,
  linkbordercolor=red,
}
\makeatletter
\Hy@AtBeginDocument{%
  \def\@pdfborder{0 0 1}
  \def\@pdfborderstyle{/S/U/W 1}
}
\makeatother
\usepackage{colortbl}
\usepackage[normalem]{ulem}
\useunder{\uline}{\ul}{}
\usepackage{lscape}
\usepackage[
backend=biber,
style=numeric,
sorting=ynt
]{biblatex}
\usepackage{caption} 
\def\BibTeX{{\rm B\kern-.05em{\sc i\kern-.025em b}\kern-.08em
    T\kern-.1667em\lower.7ex\hbox{E}\kern-.125emX}}

\title{Mapping LLM Security Landscapes: A Comprehensive Stakeholder Risk Assessment Proposal\\

\makeatletter
\newcommand{\linebreakand}{%
  \end{@IEEEauthorhalign}
  \hfill\mbox{}\par
  \mbox{}\hfill\begin{@IEEEauthorhalign}
}
\makeatother

}

\author{
    \IEEEauthorblockN{Rahul Pankajakshan\IEEEauthorrefmark{1},
    Sumitra Biswal* \IEEEauthorrefmark{2}, Yuvaraj Govindarajulu*\IEEEauthorrefmark{2}, Gilad Gressel\IEEEauthorrefmark{1}\\
    }
    \IEEEauthorblockA{\IEEEauthorrefmark{1}Center for Cybersecurity Systems and Networks, Amrita Vishwa Vidyapeetham, Amritapuri, India\\
    \{rahulp, gilad.gressel\}@am.amrita.edu}
    \IEEEauthorblockA{\IEEEauthorrefmark{2}AIShield, Bosch Global Software Technologies, Bengaluru, India\\
    \{sumitra.biswal, govindarajulu.yuvaraj\}@in.bosch.com}
    
}

\begin{document}

\maketitle

\begin{abstract}
The rapid integration of Large Language Models (LLMs) across diverse sectors has marked a transformative era, showcasing remarkable capabilities in text generation and problem-solving tasks. However, this technological advancement is accompanied by significant risks and vulnerabilities. Despite ongoing security enhancements, attackers persistently exploit these weaknesses, casting doubts on the overall trustworthiness of LLMs. Compounding the issue, organisations are deploying LLM-integrated systems without understanding the severity of potential consequences. Existing studies by OWASP and MITRE offer a general overview of threats and vulnerabilities but lack a method for directly and succinctly analysing the risks for security practitioners, developers, and key decision-makers who are working with this novel technology. To address this gap, we propose a risk assessment process using tools like the OWASP risk rating methodology which is used for traditional systems. We conduct scenario analysis to identify potential threat agents and map the dependent system components against vulnerability factors. Through this analysis, we assess the likelihood of a cyberattack. Subsequently, we conduct a thorough impact analysis to derive a comprehensive threat matrix. We also map threats against three key stakeholder groups: developers engaged in model fine-tuning, application developers utilizing third-party APIs, and end users. The proposed threat matrix provides a holistic evaluation of LLM-related risks, enabling stakeholders to make informed decisions for effective mitigation strategies. Our outlined process serves as an actionable and comprehensive tool for security practitioners, offering insights for resource management and enhancing the overall system security.

\end{abstract}

\def\thefootnote{*}\footnotetext{These authors contributed equally to this work.}\def\thefootnote{\arabic{footnote}}
\section{Introduction}

Large Language Models (LLMs) have become increasingly prevalent in various sectors, including industry and academia. They are undergoing transformative integration at an impressive pace, showcasing remarkable human-like capabilities in text generation, language understanding, sentiment analysis, and summarization. They also demonstrate appreciable abilities in code generation, problem-solving, and reasoning tasks. The accessibility and affordability of LLM inference APIs and the availability of open-source models have facilitated their widespread adoption, resulting in a continuous emergence of a plethora of applications.

However, LLMs embody a dual nature, wielding the potential to be both a powerful tool for progress and innovation and a concerning source of harm and misuse. They introduce a new set of risks, which are evolving and not understood well enough to be effectively managed. These include attacks like prompt injection to hijack the model and perform unauthorized actions, as well as planting backdoors in training data for potentially malicious activities once deployed. Despite significant efforts to align the models and implement defensive mechanisms to make LLMs more helpful and less harmful \cite{ouyang2022training, bai2022training}, attackers have found ways to circumvent these guardrails \cite{huang2023catastrophic, wolf2023fundamental}. Nevertheless, organisations and users are adopting this technology without understanding its security and privacy implications. For instance,  an employee at Samsung accidentally leaked proprietary code via ChatGPT \cite{Ray_2023}, while Amazon's recently implemented LLM-based chatbot, Q, inadvertently disclosed confidential information and generated severely hallucinatory responses \cite{SchifferQbot_2023}. This low maturity level and bug-ridden experience are cause for concern and significantly negatively impact their trustworthiness and usability in the future.

The Open Web Application Security Project (OWASP), with the help of industry and academic experts, has compiled a detailed list of the top 10 vulnerabilities and threats to LLM applications \cite{OWASPtop10}.  MITRE has introduced the ATLAS threat matrix, which outlines the adversarial tactics and techniques used to attack AI systems, based on the popular ATT\&CK framework\cite{MITREAtlas}. These studies have provided a comprehensive understanding of the various attack methods in the LLM ecosystem. However, they lack focus on identifying the severity of risks faced by different stakeholders within the LLM ecosystem. Unlike traditional IT systems, there is an absence of well-established risk assessment or threat modeling processes for LLM-based systems. This absence poses challenges for security professionals in evaluating cyber risks and implementing appropriate mitigation strategies to secure their systems effectively.




In this study, we demonstrate the applicability of tools like the OWASP Risk Rating Methodology, commonly utilized for evaluating risks in conventional IT systems, to assess the severity of risks specific to LLM-based systems. Our approach involves a three-point method: first, we conduct a scenario analysis, considering factors such as the motive and skill level of potential threat agents. Next, we map dependent system components against vulnerabilities, taking into consideration factors such as the ease of system exploitation and vulnerability discovery. By integrating scenario analysis with dependency mapping, we estimate the likelihood of a threat. The final step in our approach involves performing an impact analysis, aimed at comprehending both the technical and business implications. Following our detailed risk analysis, we distilled our findings into a threat matrix, providing stakeholders with a quick-reference tool that can be tailored to their needs. We'll demonstrate this risk assessment process on a hypothetical use case to show the usability and relevance of the process.

We argue that this risk assessment process, along with the threat matrix, would provide readers involved in the risk assessment of their LLM-based systems with a valuable and actionable tool for resource management and understanding the overall security posture of their system. Moreover, such an approach empowers security professionals and developers to make informed decisions regarding risk mitigation by gaining insights into the specific threats that may impact them and optimise resource allocation and management. 



\section{Background}
\subsection{Rise of LLMs}
LLMs represent a significant advancement in natural language processing. These deep learning algorithms, trained on massive datasets of text and code, boast capabilities far exceeding their predecessors in understanding and generating human-like text. The transformer architecture, introduced by Vaswani et al. in 2017, serves as the foundational framework for large language models, facilitating their ability to capture intricate dependencies in sequential data through self-attention mechanisms\cite{vaswani2017attention}. 

The training process of LLMs mainly involves two key phases: pretraining and fine-tuning. In the pretraining phase, the model undergoes self-supervised learning on a diverse dataset, predicting the next word in a sentence based on context. This self-supervised approach and techniques like Masked Language Modelling (MLM) and Casual Language Model (CLM) enable the model to grasp complex language patterns implicitly.  This results in a pre-trained LLM with a generalised understanding of language patterns. Subsequently, in the fine-tuning phase, the model is adapted to specific tasks or domains using smaller, task-specific datasets. Fine-tuning employs strategies like Reinforcement Learning from Human Feedback (RLHF) \cite{christiano2017deep}, where the model refines its capabilities through iterative adjustments based on human-provided feedback. This process enhances the LLM's proficiency in performing targeted applications such as text summarization, translation, or sentiment analysis, ensuring its adaptability and effectiveness in real-world scenarios.

\subsection{Risk Assessment}
Risk assessment is a crucial component in information security management, providing organisations with a systematic approach to identifying, analysing, and mitigating potential risks. Various frameworks have been established over the years to guide the risk assessment and management process, each with its unique perspective and methodologies. Some prominent frameworks include the ENISA Risk Management Framework \cite{ENISA}, the NIST Risk Management Framework \cite{NIST}, and the ISO 27001 \cite{ISO}.

However, they all follow a common approach involving the following key steps:

\begin{enumerate}
    \item \textbf{Risk Identification:} Initiate the process by identifying potential threats, vulnerabilities, and assets susceptible to security breaches.

    \item \textbf{Risk Analysis:} Analyze the identified threats using a risk rating methodology. Estimate the likelihood and impact of risks to assess their significance.  

    \item \textbf{Risk Evaluation:} Prioritize risks based on their potential impact, aligning with the organisation's objectives.

    \item \textbf{Risk Treatment:} Take action to manage the identified risks. Develop and implement strategies to mitigate, transfer, or accept risks, considering risk criticality and aligning with the organization's security policies and tolerance.

    \item \textbf{Monitoring and Review:} Continuously assess and update the risk management process to address evolving threats and vulnerabilities.

\end{enumerate}

In addition to the frameworks, risk assessments can be categorised into different types based on their methodologies:

\begin{itemize}
    \item \textbf{Qualitative Risk Assessment:} Focuses on subjective judgment and expert opinions to assess risks based on qualitative criteria such as high, medium, or low.
    \item \textbf{Quantitative Risk Assessment:} Utilises numerical data and statistical methods to quantify risks' potential impact and likelihood, providing a more precise analysis.
    \item \textbf{Semi-Quantitative Risk Assessment:} Combines elements of both qualitative and quantitative assessments to provide a balanced approach, incorporating expert judgment and numerical data.
\end{itemize}


\subsection{OWASP Top 10 for LLM}
The OWASP Top 10 for LLM Applications is a living document that serves as a comprehensive guide for developers and security teams navigating the unique security challenges posed by large language models, such as prompt injection, data poisoning, and model theft. It identifies the top ten critical vulnerabilities, details example attack scenarios, and recommends specific mitigation strategies to help developers build secure and reliable LLM applications.

\subsection{Risk Rating} \label{riskrating}
Risk rating entails estimating the likelihood of an attack occurring and, should it happen, evaluating its potential impact. Several well-established and widely recognized methodologies exist to assist in calculating both the likelihood and impact, such as NIST SP 800-30\cite{NISTSP} and the Harmonized Threat and Risk Assessment (TRA) Methodology from the Canadian Centre for Cyber Security \cite{TRA}. Alternatively, the OWASP Risk Rating Methodology\cite{OWASPRisk} offers a more straightforward and simplified approach.

The OWASP Risk Rating Methodology is a structured approach that assists organizations in evaluating and prioritizing potential risks associated with software and web application security. It adheres to a standard model for risk calculation, expressed as:

\begin{equation} \label{riskequation}
\text{Risk} = \text{Likelihood} \times \text{Impact}
\end{equation}

\subsubsection{\textbf{Likelihood factors}}
The factors applicable to estimating the likelihood of risk are categorized into two distinct groups:

\begin{itemize}
    \item \textit{Threat Agent Factors:} These factors encompass (i) the skill level of the threat actor, (ii) their motive to exploit, (iii) the opportunities or resources required for the threat actor to discover and exploit the vulnerability, and (iv) the size of the threat actor group.

    \item \textit{Vulnerability Factors:} This group includes factors such as the (i) ease of discovering the vulnerability by threat agents, (ii) the ease of exploiting the vulnerability once discovered, (iii) awareness of the vulnerability among threat agents, and the likelihood of a successful (iv) intrusion detection if the exploit occurs.
\end{itemize}

\subsubsection{\textbf{Impact factors}}

The factors used to determine impact are also categorized into two groups:
\begin{itemize}
    \item \textit{Technical Impact Factors:} These factors concentrate on the traditional cybersecurity areas of concern, namely, (i) loss of confidentiality, (ii) loss of integrity, (iii) loss of availability, and (iv) loss of accountability.

    \item \textit{Business Impact Factors:} These factors are crucial to the organisation and encompass considerations such as the (i) financial damage resulting from the exploit, the (ii) impact on the business's reputation, (iii) the exposure introduced by non-compliance, and the (iv) severity of privacy violations.
    \end{itemize}

The factors are scored on a scale from 0 to 9. The likelihood score is determined by averaging the scores of the threat agent factor and the vulnerability factors, while the impact score is derived by averaging the scores of the technical and business impact factors. The values of these factors can also be weighted according to specific requirements. The likelihood and impact levels chart, as well as the overall risk severity chart, can be configured according to the specific needs of the organization, or standard tables proposed by OWASP can be adopted.

\section{LLM Risks}
This section is dedicated to enumerating all the risks and threats associated with LLMs, adhering to the taxonomy outlined by the OWASP in their Top Ten for LLMs version 1.1.0 \cite{OWASPtop10}. This approach is employed to ensure comparability and mitigate redundancy in the identification and categorisation of potential security concerns related to LLMs.

\subsection{LLM01: Prompt Injection}

Prompt Injection is an LLM vulnerability that enables an attacker to manipulate the LLM's output by carefully crafted prompts, leading to the generation of texts that usually violate the LLM's developer-set usage policies. There exist two primary categories of prompt injections:

\begin{itemize}
    \item \textit{Direct Prompt Injections}: Colloquially known as ``jailbreaking", these entail the manipulation of the system prompt through overwriting or revealing, often leading to partial IP loss. This may involve crafting prompts with the specific intent of circumventing safety and moderation features imposed on LLMs by their creators. 

    \item \textit{Indirect Prompt Injections}: Occurs when an LLM accepts input from external sources susceptible to control by an attacker, such as websites or files. In this context, attackers can deceive the LLM into interpreting input from LLM as ``commands" rather than ``data" for processing, consequently inducing unexpected behaviour in LLM-based applications or compromising the security of the entire system \cite{greshake2023more}.
\end{itemize}

Automated tools for jailbreaking LLMs have been developed\cite{deng2023jailbreaker}, as well as multi-prompt injection techniques\cite{li2023multi} are discussed in the existing literature. Moreover, universal and transferable adversarial suffixes have emerged as effective methods for jailbreaking various models \cite{zou2022attacks}.

\subsection{LLM02: Insecure Output Handling}
General-purpose LLMs undergo training on a substantial portion of the internet. When employed for downstream tasks in any application or plug-ins, developers must exercise caution in their utilisation, as these models can generate outputs that may be harmful to the user or the application itself. Insecure Output Handling specifically pertains to the absence of sufficient validation or sanitisation of LLM outputs before they are used for downstream tasks. If the outputs of LLMs are not managed properly, it could lead to security risks like Cross-Site Scripting and Cross-Site Request Forgery in web browsers. Attackers can also exploit the LLM outputs for privilege escalation, and remote code execution on backend systems \cite{MathGPT}.

\subsection{LLM03: Training Data Poisoning}
Training data poisoning involves deliberately manipulating the data used to train these models with malicious intent. Adversaries strategically inject deceptive or biased examples into the training dataset at the pre-training or fine-tuning stage, aiming to influence the model's learning process. Attackers can introduce backdoors, biases or other vulnerabilities that can degrade the model's security, performance, and trustworthiness \cite{xu2023instructions}.

\subsection{LLM04: Model Denial of Service}
LLMs are very resource-intensive to train and run. An attacker can interact with LLMs leading it to consume resources excessively resulting in a decline in the quality of service or even denial of service to other users as well as higher compute costs. Attackers can craft prompts that are computationally complex in terms of context length or language patterns.

\subsection{LLM05: Supply Chain Vulnerabilities}
In the context of LLMs, the supply chain refers to the entire process from data collection and model training to deployment. It involves various components such as the training data, pre-trained models, and the deployment infrastructure. Each component can be vulnerable, the crowd-sourced training data could be poisoned, the pre-trained model could be compromised or the third-party packages used to develop the LLM could be insecure.

\subsection{LLM06: Sensitive Information Disclosure}
LLMs are pre-trained on diverse datasets that include snippets of real-world data. During the generation process, these models can inadvertently produce responses that disclose sensitive details. Conversational agents like OpenAI's ChatGPT and Google's Gemini collect user prompts during conversations to enhance their model's performance. However, this practice introduces a security and privacy concern, as the model may unintentionally generate outputs that reveal confidential or private information. Moreover, using carefully crafted prompts, an attacker could exploit this vulnerability to reveal or expose sensitive details intentionally.

\subsection{LLM07: Insecure Plugin Design}
Often LLM plugins accept user input as free text, which can be easily exploited by an attacker. The LLM plugins that are designed without proper access control or input validation can result in SQL injection or remote code execution.

\subsection{LLM08: Excessive Agency}

LLM-based systems make decisions based on a user's prompt or the input they receive from another integrated component. If the degree of freedom or authorisation granted to the LLM is excessive, attackers can exploit this vulnerability to compromise the LLM-based system. However, an attacker need not exploit this vulnerability to be harmful. Any unsuspecting user input or unintended action from a system component can lead the model to produce ambiguous or unexpected output, causing the system to behave unexpectedly. For instance, an LLM-based file summarizer utilizes a third-party plugin for reading files from the user. However, this plugin also possesses the capability to modify and delete files. If a user detects an error in the LLM's generated response, they may report the mistake to the application, directing the LLM to potentially modify or delete the files \cite{OWASPtop10}.

\subsection{LLM09: Overreliance}
LLMs can ``hallucinate", generating information that can be factually incorrect, unsafe, or inappropriate \cite{gallegos2023bias, zhang2023siren}. When these models are relied upon to generate source code, there is a risk of introducing unnoticed security vulnerabilities, which pose a significant threat to the safety and security of applications as well its users. Relying on such information or code without adequate oversight can result in security breaches, spread of misinformation, communication breakdowns, legal complications, and damage to one's reputation.

\subsection{LLM10: Model Theft}
Model theft is the illegal act of copying or extracting weights or parameters or data from closed-source LLM models to create functional equivalents\cite{taori2023alpaca}. This activity can lead to substantial economic losses and harm to brand reputation, posing a threat to competitive advantage. Attackers may exploit the proprietary information within the model or use the model itself for malicious purposes.

\section{Methodology}

In this section, we will outline the methodology employed to derive the threat matrix. We will begin by discussing the scope of the identified stakeholder groups. Following that, we will delve into the risk analysis process that we propose, which is rooted in the OWASP risk rating methodology. This risk assessment process underpins our threat matrix, ensuring each risk is systematically analyzed and rated.

\subsection{Stakeholders to LLM}\label{stakeholders}
The stakeholders involved in large language models range from prominent corporations funding and advancing these models to individuals within the broader populace who might be unaware of their existence. This paper focuses on stakeholders directly affected by potential security lapses in large language models. We specifically highlight three key groups of stakeholders: developers and organisations engaged in fine-tuning open-source pre-trained models such as  Llama-2 family \cite{touvron2023llama} or Mistral family\cite{jiang2023mistral} for various downstream tasks, application developers who depend on third-party LLM inference APIs such as GPT-3.5, GPT-4 provided by OpenAI, and end users who utilise these systems.

\subsubsection{\textbf{LLM Fine-tuning Developers}} 

The significant financial investment required for developing foundational models from scratch includes expenditures on cutting-edge computing infrastructure for model training, such as thousands of powerful GPUs, extensive datasets, and the recruitment of specialized research and development teams. Unfortunately, this level of investment is feasible for only a limited number of organisations. This significant financial barrier presents a formidable challenge for smaller organizations, limiting their capacity to compete on an equal footing.

Despite these challenges, open-source large language model families, comprising both pre-trained and fine-tuned models, are increasingly becoming accessible for public and commercial use \cite{touvron2023llama, jiang2023mistral}. This accessibility creates an opportunity for less resource-constrained entities in LLM research and development to fine-tune their versions for various tasks. Techniques like Low-Rank Adaptation (LoRA) \cite{hu2021lora} have also helped preserve performance while reducing memory requirements and accelerating training. Fine-tuning typically involves data collection, establishing training infrastructure, and leveraging expertise in machine learning. Through this process, the model's knowledge becomes refined and tailored to the specific domain of the application, resulting in more relevant and nuanced outputs.

\subsubsection{\textbf{LLM API Integration Developers}}

The emergence of ChatGPT and its rapid rise in popularity has spurred the development of applications and plugins based on LLMs. These applications primarily utilize public LLM inference APIs. These APIs allow developers to send input data or prompts to the LLM and receive its output without the need to manage the model or its training process directly. While this approach offers convenience, developers rely on the API provider for maintaining and enhancing the underlying LLM, limiting their ability to customize its behaviour. 

These applications span various domains, such as content creation assistants, coding or problem-solving tools, and even support for HR recruitment processes. OpenAI has further facilitated this trend through the introduction of the `GPT store', creating a platform for the launch of millions of custom versions of ChatGPT or simply `GPTs'.

However, in the rush to adopt cutting-edge technology, there is a significant risk that security considerations are being overlooked, either due to cost constraints or a lack of awareness, given the newness of this technology. For instance, MathGPT is an LLM-based application intended to aid users in solving mathematical problems. A user was able to inadvertently expose its OpenAI GPT-3.5 API key through code execution, exploiting a vulnerability related to prompt injection \cite{MathGPT}.

\subsubsection{\textbf{End Users}}

End users constitute a pivotal stakeholder group in the domain of LLM. They are the ultimate consumers of applications, services, or platforms powered by LLMs, interacting with technology that influences various aspects of their lives, including education, communication, business, and entertainment. Users depend on LLM-based applications to receive assistance in writing, access health-related information, and make informed decisions in finance. These models have become integral to contemporary life, permeating multiple facets and delivering convenience and efficiency across diverse walks of life. It is crucial to prioritise the safety, privacy, and security of these end users to maintain their trust and uphold a positive outlook towards large language models.
 
\subsection{Risk analysis process} \label{riskanalysis_process}
The risk analysis process starts with the initial step of risk identification, wherein all potential threats are enumerated based on the taxonomy proposed by OWASP in their Top Ten for LLM.

In assessing each identified risk, we utilize both the risk factors and recommendations outlined in the OWASP Risk Rating Methodology to determine its criticality. This process can be characterized as a semi-quantitative risk assessment. The overall risk assessment process we advocate adopts a three-step approach to determine the values for likelihood and impact factors:

\begin{enumerate}
    \item \textbf{Scenario Analysis:} In this phase, targeted scenarios are crafted, and the rationale behind threat agent factors is deliberated upon. The strategy we are following in this step involves identifying the threat agents who are most likely to exploit the vulnerability and examining the worst-case scenario.
    
    \item \textbf{Dependency Mapping:} All system components associated with the vulnerability are identified and mapped against the vulnerability factors, such as ease of discovery, ease of exploitation, awareness of the threat agent regarding the vulnerability, and the likelihood of detecting the intrusion when the exploit occurs. The values for vulnerability factors need to be estimated based on the threat agent factors. Once a clear understanding of threat agents and vulnerability factors is obtained, the likelihood of the attacker targeting the vulnerability is estimated using the standard risk calculation equation (\ref{riskequation}).

     \item \textbf{Impact Analysis:} This crucial step involves assessing the technical and business impact if the exploit occurs. The technical impact is initially assessed by contemplating worst-case scenarios, providing a foundation for estimating the subsequent impact on the business operations.
\end{enumerate}


\section{Threat Matrix }
In this section, we present a generic threat matrix in \autoref{tab:threat-matrix}. We map the OWASP Top Ten LLM-specific threats against different stakeholders, providing practitioners with a reference sheet for conducting risk assessments. This matrix equips stakeholders with the insights needed for targeted risk mitigation, summarising the comprehensive analysis in an accessible format.

In this threat matrix, we have outlined the general causes and consequences of each LLM-relevant risk, along with corresponding suggested controls and mitigation, after conducting an extensive review of existing literature. We also indicate whether the risk falls within the realm of traditional cybersecurity risks. For instance, as mentioned in the matrix, the vulnerability `Prompt Injection' typically is caused by insufficient control over input and the LLM's inherent nature to assist users with their queries. The potential consequences of exploiting this vulnerability vary from reputation harm to user harm through indirect prompt injection. Mitigation techniques that may be utilized include static input validation and dynamic output filtering. Since these vulnerabilities have emerged with the advent of LLMs and other foundational models, it is placed outside the scope of traditional cybersecurity risks. However, this risk affects all identified stakeholder groups we are targeting.

The probability, impact, and risk rating values within the matrix have deliberately been left blank to maintain generality and serve their purpose as a template. We will now demonstrate how to calculate the values for the probability and impact field using a hypothetical scenario for the `LLM fine-tuning developers' stakeholder group. This process will aid in deriving the final risk rating.

\begin{table*}[]
\centering
\resizebox{\textwidth}{!}{%
\begin{tabular}{@{}lllllllll@{}}
\toprule
\textbf{S.No} &
  \textbf{Risk   Description} &
  \textbf{Cause \&   Consequences} &
  \textbf{Likelihood} &
  \textbf{Impact} &
  \textbf{\begin{tabular}[c]{@{}l@{}}Risk \\ Rating\end{tabular}} &
  \textbf{Controls/Mitigation} &
  \textbf{\begin{tabular}[c]{@{}l@{}}Traditional  \\  Cybersec\end{tabular}} &
  \textbf{Concerned Stakeholders} \\ \midrule
LLM01 &
  \begin{tabular}[c]{@{}l@{}}Prompt Injection \end{tabular} &
  \begin{tabular}[c]{@{}l@{}}Caused by: lack of   control/validation on\\      LLM's input, LLM's implicit nature or \\      design/architecture \\      \\      Consequences:   Reputation loss, Partial IP loss, \\      Performance degradation, User harm\end{tabular} &
   &
   &
   &
  \begin{tabular}[c]{@{}l@{}}Static: Use trusted/reputed LLM   service \\      provider, Input validation and filtering\\      \\      Dynamic: Adaptive   trust boundaries for \\      input source, Monitoring of LLM outputs, Red \\ teaming, LLM Response monitoring/filtering\end{tabular} &
  No &
  \begin{tabular}[c]{@{}l@{}}LLM   Fine-tuning Developers\\      LLM API Integration Developers\\      End users\end{tabular} \\ \midrule
LLM02 &
  \begin{tabular}[c]{@{}l@{}}Insecure  \\  Plugin Design\end{tabular} &
  \begin{tabular}[c]{@{}l@{}}Caused by: improper access control, poor \\      design and implementation\\      \\      Consequences:   Compromised System\end{tabular} &
   &
   &
   &
  \begin{tabular}[c]{@{}l@{}}Static: Input sanitisation,   parameterisation, \\      validation, protect against all REST API security\\      risks\\      \\      Dynamic: proper authorisation and authentication\end{tabular} &
  Yes &
  \begin{tabular}[c]{@{}l@{}}LLM   Fine-tuning Developers\\      LLM API Integration Developers\end{tabular} \\ \midrule
LLM03 &
  \begin{tabular}[c]{@{}l@{}}Training  \\  Data Poisoning\end{tabular} &
  \begin{tabular}[c]{@{}l@{}}Caused by: Poor vetting/verification of training \\      data and data source\\      \\      Consequences:   Reputation loss, Model integrity \\      loss, Financial Damage, Misinformation and \\      bias, Performance degradation, User harm\end{tabular} &
   &
   &
   &
  \begin{tabular}[c]{@{}l@{}}Static: Exhaustive analysis and   sanitisation of all \\      unvetted training dataset.\end{tabular} &
  No &
  \begin{tabular}[c]{@{}l@{}}LLM   Fine-tuning Developers\\      End users\end{tabular} \\ \midrule
LLM04 &
  \begin{tabular}[c]{@{}l@{}}Model   Denial \\ of Service\end{tabular} &
  \begin{tabular}[c]{@{}l@{}}Caused by: Poor design and implementation, \\      improper input validation\\      \\      Consequences:   Financial and Reputation loss\end{tabular} &
   &
   &
   &
  \begin{tabular}[c]{@{}l@{}}Static: Use proper input validation   and filtering,\\      rate-limiting, usage limit per user, Adversarial \\      input detection\\      \\      Dynamic: resource   utilisation monitoring\end{tabular} &
  Yes &
  \begin{tabular}[c]{@{}l@{}}LLM   Fine-tuning Developers\\      LLM API Integration Developers\\      End users\end{tabular} \\ \midrule
LLM05 &
  \begin{tabular}[c]{@{}l@{}}Supply   Chain \\ Vulnerabilities\end{tabular} &
  \begin{tabular}[c]{@{}l@{}}Caused by: Poor security review and vetting of\\      3rd party components used\\      \\      Consequences:   Variable - Compromised System,\\       Performance degradation\end{tabular} &
   &
   &
   &
  \begin{tabular}[c]{@{}l@{}}Static: Use only trusted/reputed 3rd   party \\      softwares and components.\end{tabular} &
  Yes &
  \begin{tabular}[c]{@{}l@{}}LLM   Fine-tuning Developers\\      LLM API Integration Developers\end{tabular} \\ \midrule
LLM06 &
  \begin{tabular}[c]{@{}l@{}}Sensitive   Information \\ Disclosure\end{tabular} &
  \begin{tabular}[c]{@{}l@{}}Caused   by: Incomplete training data sanitization, \\      training data memorisation                                           \\      Consequences: Privacy violation, Reputation  \\  damage, Partial IP loss, User harm\end{tabular} &
   &
   &
   &
  \begin{tabular}[c]{@{}l@{}}Static: Training data monitoring to weed out\\      sensitive information, Differential privacy \\      mechanisms, Encrypt sensitive information     \\      Dynamic:\end{tabular} &
  No &
  \begin{tabular}[c]{@{}l@{}}LLM   Fine-tuning Developers\\      LLM API Integration Developers\\      End users\end{tabular} \\ \midrule
LLM07 &
  \begin{tabular}[c]{@{}l@{}}Insecure   Output \\ Handling\end{tabular} &
  \begin{tabular}[c]{@{}l@{}}Caused by: General purpose LLM's ability to \\      generate arbitrary code and text, improper \\      input validation or output scrutiny\\      \\      Consequences: IP   loss, Compromised system \\      and data, User harm\end{tabular} &
   &
   &
   &
  \begin{tabular}[c]{@{}l@{}}Static: Proper validation/filtering   of output, \\      output encoding to mitigate code execution,\\  rate limiting\end{tabular} &
  No &
  \begin{tabular}[c]{@{}l@{}}LLM   Fine-tuning Developers\\      LLM API Integration Developers\\      End users\end{tabular} \\ \midrule
LLM08 &
  Excessive Agency &
  \begin{tabular}[c]{@{}l@{}}Caused by: design and implementation choices,\\      improper access control\\      \\      Consequences:   Variable - Compromised System\end{tabular} &
   &
   &
   &
  \begin{tabular}[c]{@{}l@{}}Static: Limit the permissions of   LLMs , \\      use components with granular functionalities \\      than open-ended ones,\\      \\      Dynamic: implement  proper authorisation\end{tabular} &
  No &
  \begin{tabular}[c]{@{}l@{}}LLM   Fine-tuning Developers\\      LLM API Integration Developers\\      End users\end{tabular} \\ \midrule
LLM09 &
  Overreliance &
  \begin{tabular}[c]{@{}l@{}}Caused by: blindly trusting LLM generated content\\ without review\\ \\ Consequences: Misinformation, implementation of \\ incorrect solutions\end{tabular} &
   &
   &
   &
  \begin{tabular}[c]{@{}l@{}}Static: User awareness\\ \\ \\ Dynamic: Output validation and review,\end{tabular} &
  No &
  End Users \\ \midrule
LLM10 &
  Model   Theft &
  \begin{tabular}[c]{@{}l@{}}Caused   by: Weak access control, insider threats, \\      model inversion\\           Consequences: Reputation loss, Model integrity   \\      loss, Financial Damage, Misinformation, privacy \\      violation\end{tabular} &
   &
   &
   &
  \begin{tabular}[c]{@{}l@{}}Static:   Model obfuscation     \\      Dynamic: Strong access controls and  \\  authentication, Regular auditing\end{tabular} &
  No &
  LLM   Fine-tuning Developers \\ \bottomrule
\end{tabular}%
}
\caption{
The generic threat matrix for LLM enumerates the risks identified in OWASP's Top Ten for LLM, detailing causes, consequences, and control measures, both static and dynamic. Each risk is also categorized as a traditional cyber risk or not. Likelihood, impact, and risk ratings should be calculated using the proposed assessment process.  Furthermore, the matrix maps these risks against the respective stakeholder groups, facilitating easy reference and analysis. }
\label{tab:threat-matrix}
\end{table*}

\section{Use case Analysis}\label{usecaseSection}
 We will conduct a risk assessment for a university virtual assistant. This virtual assistant is created by fine-tuning an open-source pre-trained LLM using a dataset containing course and administrative information from the university. 


\subsection{University Virtual Assistant}\label{usecaseExampleA}
The University Virtual Assistant is a chatbot developed by a prominent university to help students and faculty with details regarding the course materials, university policies, and other administrative information. This assistant is developed by fine-tuning an open-source pre-trained LLM with databases containing lecture materials from courses with a high number of attendees, course details, and administrative instructions from the university. The university has developed the bot to enhance administrative efficiency by automating routine tasks and freeing up staff resources. Also, the university aims to improve the student experience by offering personalized assistance with coursework and ensuring availability 24/7.

\subsection{System Design and Analysis}

\subsubsection{System Description} Authenticated students and faculty of the university can utilise the virtual assistant to inquire about course details, requirements, and lecture materials. It also offers information regarding campus resources, services, administrative procedures, policies, and general details about the university and its surroundings.

\begin{figure*}
    \centering
    \includegraphics[width=0.8\textwidth]{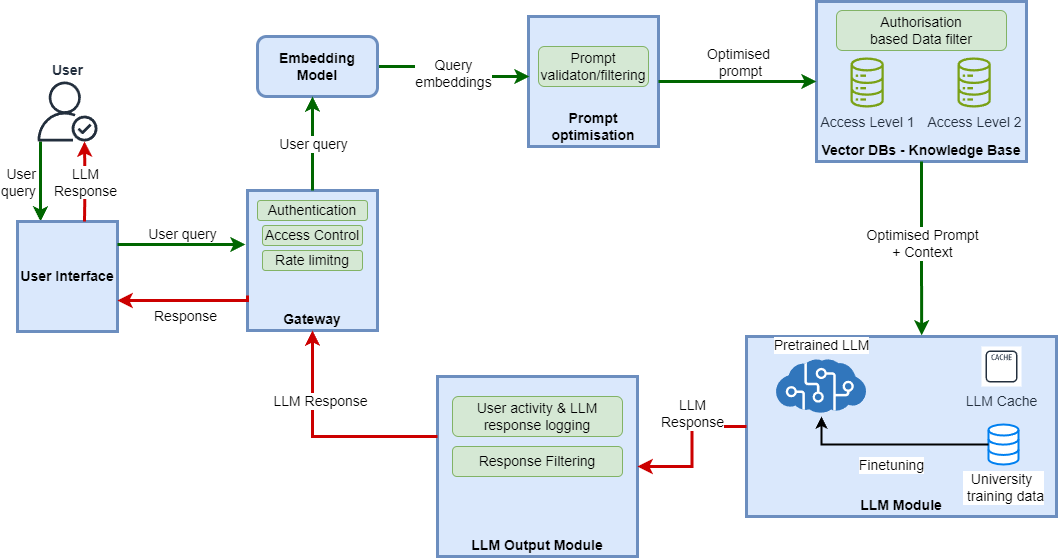}
    \caption{The system design of the university virtual assistant emphasizes the diverse security measures implemented for each component. An authenticated user's prompt undergoes optimization and validation/filtering for prompt injection within the prompt optimization module. Contextual information for the query is accessed based on authorization. The query and context serve as inputs to the LLM, and the resulting response from the LLM is logged and cached. Before being sent to the user, the response undergoes filtering. A rate-limiting mechanism has been implemented at the Gateway to safeguard against DoS attacks.}
    \label{fig:usecaseA_arch}
\end{figure*}

The pre-trained model is fine-tuned using a dataset comprising student and faculty-related questions with human-reviewed responses. Additionally, the model has access to a knowledge base that is regularly updated with information from the university to assist in answering queries. Access levels to the knowledge base are differentiated, with students having access only to information from level 1 for their specific requirements. On the other hand, Access Level 2 contains sensitive information intended for use as contextual data by faculty and administrative staff, such as income from students or PII of students and staff, such as gender and race. \autoref{fig:usecaseA_arch} illustrates the system design of the University Virtual Assistant with the security components highlighted.

\subsubsection{Security Overview} The system is designed with a focus on usability and accessibility, with an overall security posture of moderate strength. The system includes an LLM user input validation and filtering library that has not undergone rigorous testing for vulnerabilities such as role-promoting injection, which threat actors with this knowledge could exploit. The library performs a similarity search using a set of historical prompt injection attacks or attempts. However, it lacks the capability to detect any variations. The system lacks multi-factor authentication and has a modest password policy that can be vulnerable to brute-force attacks.

The training dataset used for fine-tuning the model is benchmarked using only a single dataset for detecting obvious bias and toxicity; there is no extensive vetting or verification process with multiple benchmark datasets in place during the fine-tuning of the model. While data sources are audited manually, there is no real-time monitoring for malicious activity or anomaly detection. The system relies on well-established libraries and packages but does not update them automatically with available patches. Basic rate limiting with adjustable thresholds is enforced to prevent DoS attacks.

User activities are logged, but advanced statistical analysis in real-time to detect suspicious activities is not performed. User awareness campaigns on basic security and system functionalities are conducted. In short, we imagine a university with limited resources making the best effort - the maturity level of security operations is slightly naive combined with a lack of resources.


\begin{table*}[]
\centering
\resizebox{\textwidth}{!}{%
\begin{tabular}{@{}llll@{}}
\toprule
\multicolumn{1}{c}{\textbf{\begin{tabular}[c]{@{}c@{}}Vulnerability\\  Description\end{tabular}}} &
  \multicolumn{1}{c}{\textbf{Scenario analysis}} &
  \multicolumn{1}{c}{\textbf{Dependency mapping}} &
  \multicolumn{1}{c}{\textbf{Impact Analysis}} \\ \midrule
\begin{tabular}[c]{@{}l@{}}Prompt Injection \\ \end{tabular} &
  \begin{tabular}[c]{@{}l@{}}Threat Agent: An authenticated student \\ or someone who has bypassed authentication.\\ \\ Skill Level: Network and programming skills\\  - The potential threat agent or a university student is \\ expected to have good programming skills and \\ knowledge of LLM.\\ \\ Motive: Possible rewards \\ - Threat agent might be interested in academic exploits\\  or as a learning experiment.\\ \\ Opportunity: Some access or resources required \\ - Basic user account access required.    \\ \\ Size: Authenticated users \\ - Any of the authenticated student or attackers \\ who has bypassed authentication can be the attacker.\\ \\ Method: Crafts malicious prompts\end{tabular} &
  \begin{tabular}[c]{@{}l@{}}Dependent Components:\\      1) Prompt validation/filtering mechanism \\           - vulnerable to role prompting\\      2) Access controls \\           - weak password policy \\      3) LLM's inherent nature \\           -LLM's are aligned to be helpful, engaging and responsive \\      4) LLM output/user activity monitoring \\ - logged without review\\      \\ Ease of exploit: Easy  \\ - Component 1,2, and 3 makes the exploit easy\\ \\ Ease of discovery: Automated tools available\\  - Tools available to discover the existence of prompt injection. \\ Enabled by components 1,2, and 3.   \\ \\ Awareness: Public knowledge\\  - It is public knowledge that there is no currently a robust defence \\ against prompt injection attacks.\\ \\ Intrusion Detection: Logged without review \\ - Enabled by component 4.\end{tabular} &
  \begin{tabular}[c]{@{}l@{}}Technical Impact: \\ Loss of confidentiality: Minimal non-sensitive data disclosed \\ - System prompt leaking, Access unauthorised information\\ \\ Loss of accountability: Possibly traceable \\ -  since output logs are maintained, it is possible to trace the \\ user account responsible for the exploit.\\      \\ Business Impact:   \\ Privacy Violation: Thousands of people \\ - Information   about thousands of students could be extracted.\\ \\ Non-Compliance: Clear  violation \\ -   Major privacy violation \\ \\ Financial Damage: Significant effect on annual   profit \\ - Indirect financial damage could arrive in   the form of paying \\ fines for compliance issues and privacy violations.\\ \\ Reputation Damage: Loss of goodwill \\ - Major damage to reputation possible due and people will lose  \\  trust in the application.\end{tabular} \\ \midrule
\begin{tabular}[c]{@{}l@{}}Training Data \\ Poisoning\end{tabular} &
  \begin{tabular}[c]{@{}l@{}}Threat  Agent: A malicious actor with access to \\ a popular training/finetuning dataset.\\ \\ Skill Level:  Network and programming skills\\  - Attacker should have in-depth knowledge of LLM's \\ architecture and training process to poison the training \\ data as  well as sophisticated social engineering skills to \\ smuggle the poisoned training data into the target system.\\ \\ Motive: Possible rewards \\ - Threat agent   might be a disgruntled student interested\\  in reputational damage or performance degradation or \\ trying to spread misinformation or bias. The exploit will \\ only materialise if the system incorporates the poisoned \\ data into their model.\\ \\ Size: Anonymous Internet users \\ - The attacker could be any internet user as there are possibly \\ multiple instances of poisoned data available in the internet.     \\ \\ Method: Threat agent uploads a poisoned version of a \\ popular training dataset on HuggingFace.\end{tabular} &
  \begin{tabular}[c]{@{}l@{}}Dependent   Components:\\      1) Data sources\\  - vetting or auditing of the data source is manually, so   \\ infiltration of a poisoned dataset into the system is unlikely, however, \\ they remain susceptible to human error.\\      2) LLM output/user activity monitoring \\ -    logged without review\\      3) Access logs auditing \\ - Access logs are maintained and analysed regularly for unauthorised \\        or anomalous behaviour.\\      \\ Ease of exploit: Difficult \\ - Component   1 makes it difficult to exploit.   \\  \\ Ease of discovery: Difficult \\ - Depends on   poisoned data infiltrating the system. Component 1 \\ and 3 makes it difficult. \\ \\ Awareness: Hidden \\ - Threat agents are aware of the existence of the vulnerability, \\ but it is not known whether the target system is vulnerable \\ because of the security practice of component 1.  \\ \\ Intrusion Detection: Logged without review \\ -   If the exploit happens,  it is difficult to \\ detect without active LLM output monitoring. Enabled by\\ component 2.\end{tabular} &
  \begin{tabular}[c]{@{}l@{}}Technical   Impact:\\ Loss of integrity: Extensively seriously corrupt data \\ -   Training data poisoning usually involves injecting bias or \\ misinformation in  the training data in large quantity.    \\  \\ Loss of accountability: Completely anonymous \\ -   As the threat agent has sophisticated skills to poison training \\ data, it is very likely they'll remain anonymous when uploading \\ the data for public use.     \\ \\ Business Impact:\\ Financial Damage: Significant effect on annual profit \\ - Model training/finetuning is costly. If the model is finetuned on \\ poisoned data, it cannot be fixed with continual learning. Financial \\ damage could also arrive in the form of compliance violation dues. \\     \\ Privacy Violation: Thousands of people \\ - Poisoned   model also can contain backdoors which will potentially \\ override guardrails.  Student/faculty information can be extracted. \\     \\ Non-Compliance: Clear violation - Major privacy  violation.\\      \\ Reputation Damage: Brand damage - The model   becomes completely \\ compromised. An attack of this nature results in brand damage.\end{tabular} \\ \bottomrule
\end{tabular}%
}
\caption{Risk analysis for the University Virtual Assistant use case detailing an assessment of Prompt Injection and Training Data Poisoning vulnerabilities. This analysis involves scenario analysis to estimate threat agent factors, dependency mapping to align system components with vulnerability factors, and impact analysis with documented rationale.}
\label{tab:uni-analysis}
\end{table*}

\begin{table*}[]
\centering
\resizebox{\textwidth}{!}{%
\begin{tabular}{lllll}
\hline
\textbf{} &
  \multicolumn{2}{c}{\textbf{Prompt Injection}} &
  \multicolumn{2}{c}{\textbf{Training Data Poisoning}} \\ \hline
\textbf{} &
  \textbf{Skill   level} &
  6 - Network and programming skills &
  \textbf{Skill   level} &
  6 -   Network and programming skills \\
 &
  \textbf{Motive} &
  4   - Possible reward &
  \textbf{Motive} &
  4 -   Possible reward \\
\textbf{Threat Agent Factors} &
  \textbf{Opportunity} &
  7   - Some access or resources required &
  \textbf{Opportunity} &
  0 - Full  access or expensive resources required \\
 &
  \textbf{Size} &
  6   - Authenticated users &
  \textbf{Size} &
  9 -   Anonymous Internet users \\ \hline
\textbf{} &
  \textbf{Ease of discovery} &
  9   - Automated tools available &
  \textbf{Ease of discovery} &
  3 -   Difficult \\
 &
  \textbf{Ease of exploit} &
  5   - Easy &
  \textbf{Ease of exploit} &
  3 -   Difficult \\
\textbf{Vulnerability Factors} &
  \textbf{Awareness} &
  9   - Public knowledge &
  \textbf{Awareness} &
  1 -   Unknown \\
 &
  \textbf{Intrusion   detection} &
  8   - Logged without review &
  \textbf{Intrusion   detection} &
  8 -   Logged without review \\ \hline
 &
  \textbf{Likelihood Score:} &
  \multicolumn{1}{c}{\textbf{6.75}} &
  \textbf{Likelihood Score:} &
  \multicolumn{1}{c}{\textbf{4.25}} \\
 &
  \textbf{Likelihood:} &
  \multicolumn{1}{c}{\textbf{High}} &
  \textbf{Likelihood:} &
  \multicolumn{1}{c}{\textbf{Medium}} \\ \hline
 &
   &
  \textbf{} &
  \textbf{} &
  \textbf{} \\
\textbf{} &
  \textbf{Loss of   Confidentiality} &
  5   - Extensive critical data disclosed &
  \textbf{Loss of   Confidentiality} &
  0 \\
\textbf{} &
  \textbf{Loss of Integrity} &
  0 &
  \textbf{Loss of Integrity} &
  7 -   Extensive seriously corrupt data \\
\textbf{Technical Impact Factors} &
  \textbf{Loss of   Availability} &
  0 &
  \textbf{Loss of   Availability} &
  0 \\
 &
  \textbf{Loss of   Accountability} &
  7   - Possibly traceable &
  \textbf{Loss of   Accountability} &
  9 -   Completely anonymous \\
 &
  \textbf{Technical Impact Score:} &
  \multicolumn{1}{c}{\textbf{3}} &
  \textbf{Technical Impact Score:} &
  \multicolumn{1}{c}{\textbf{4}} \\ \hline
\textbf{} &
   &
   &
   &
   \\
\textbf{} &
  \textbf{Financial   damage} &
  7 -   Significant effect on annual profit &
  \textbf{Financial   damage} &
  7 -   Significant effect on annual profit \\
 &
  \textbf{Reputation   damage} &
  5 - Loss   of goodwill &
  \textbf{Reputation   damage} &
  9 -   Brand damage \\
\textbf{Business Impact Factors} &
  \textbf{Non-compliance} &
  5 -   Clear violation &
  \textbf{Non-compliance} &
  5 -   Clear violation \\
 &
  \textbf{Privacy   violation} &
  7 -   Thousands of people &
  \textbf{Privacy   violation} &
  7 -   Thousands of people \\
 &
  \textbf{Business Impact Score:} &
  \multicolumn{1}{c}{\textbf{6}} &
  \textbf{Business Impact Score:} &
  \multicolumn{1}{c}{\textbf{7}} \\ \hline
 &
  \textbf{Final Impact Score:} &
  \multicolumn{1}{c}{\textbf{4.5}} &
  \textbf{Final Impact Score:} &
  \multicolumn{1}{c}{\textbf{5.5}} \\
 &
  \textbf{Impact:} &
  \multicolumn{1}{c}{\textbf{Medium}} &
  \textbf{Impact:} &
  \multicolumn{1}{c}{\textbf{Medium}} \\ \hline
 &
   &
   &
   &
   \\
\textbf{} &
  \textbf{Risk Severity:} &
  \multicolumn{1}{c}{\cellcolor[HTML]{FD6864}\textbf{HIGH}} &
  \textbf{Risk Severity:} &
  \multicolumn{1}{c}{\cellcolor[HTML]{FFCC67}\textbf{MEDIUM}} \\ \hline
\end{tabular}%
}
\caption{Calculating the Risk Rating using the OWASP Risk Rating Methodology for the University Virtual Assistant use case. The analysis reveals a high-risk rating for `Prompt Injection' and a medium-risk rating for `Training Data Poisoning'.}
\label{tab:uni-owasp}
\end{table*}
\subsection{Risk Analysis}

In this subsection, we will demonstrate the process of conducting risk analysis as discussed in section \ref{riskanalysis_process} for two risks: prompt injection and training data poisoning. \autoref{tab:uni-analysis} presents a detailed breakdown, including threat agent-specific scenario analysis, dependency mapping of affected system components and processes, and an assessment of the technical and business impacts. Subsequently, leveraging this analysis we determine likelihood and impact values and calculate the overall risk rating as outlined in \autoref{tab:uni-owasp} referring to the standard likelihood and impact chart, as well as the overall risk severity chart from OWASP risk rating methodology.

Our analysis of `Prompt Injection' has revealed a high likelihood of occurrence due to the ease of discovering and exploiting the vulnerability, coupled with its status as public knowledge. Additionally, we have identified a moderate impact on the business and system, primarily stemming from potential financial and reputation damage, as well as the risk of confidential information leakage.

Similarly, for `Training Data Poisoning', our analysis indicates that the likelihood of occurrence is influenced by the fact that anonymous internet users can perpetrate poisoned datasets and detecting such poisoning is challenging even after it has occurred. The impact of this risk is determined to be moderate, as it could compromise the model, resulting in financial losses and damage to the brand's reputation, given the compromised privacy of users. The threat matrix reflects these findings, assigning a high-risk rating to `Prompt Injection' and a medium rating for `Training Data Poisoning', as illustrated in \autoref{tab:threat-matrix-uni}.

\begin{table}[]
\centering
\resizebox{\columnwidth}{!}{%
\begin{tabular}{lllllll}
\hline
\textbf{S.No} &
  \textbf{Risk   Description} &
  \textbf{Cause \&   Consequences} &
  \textbf{Likelihood} &
  \textbf{Impact} &
  \textbf{\begin{tabular}[c]{@{}l@{}}Risk \\ Rating\end{tabular}} &
  \textbf{Controls/Mitigation} \\ \hline
LLM01 &
  \begin{tabular}[c]{@{}l@{}}Prompt Injection \\ and Jailbreaking\end{tabular} &
  \begin{tabular}[c]{@{}l@{}}Caused by:\\ lack of  control/validation \\ on LLM's input, \\ LLM's implicit nature or \\ design/architecture \\      \\  Consequences:  \\ Reputation loss, \\ Partial IP loss, \\ Performance degradation, \\ User harm\end{tabular} &
  \cellcolor[HTML]{FD6864}High &
  \cellcolor[HTML]{FFCC67}Medum &
  \cellcolor[HTML]{FD6864}\textbf{High} &
  \begin{tabular}[c]{@{}l@{}}Static: \\ Use trusted/reputed \\ pre-trained models,\\ Robust input \\ validation and filtering\\      \\ Dynamic: \\ Adaptive trust \\ boundaries for input \\ source, \\ Monitoring of LLM \\ outputs,\\ Red teaming, \\ LLM Response\\  monitoring/filtering\end{tabular} \\ \hline
LLM03 &
  \begin{tabular}[c]{@{}l@{}}Training  \\  Data Poisoning\end{tabular} &
  \begin{tabular}[c]{@{}l@{}}Caused by: \\ Poor vetting/verification\\ of training data and data \\ source\\      \\ Consequences:   \\ Reputation loss, \\ Model integrity loss, \\ Financial Damage, \\ Misinformation and bias, \\ Performance degradation, \\ User harm\end{tabular} &
  \cellcolor[HTML]{FFCC67}Medium &
  \cellcolor[HTML]{FFCC67}Medium &
  \cellcolor[HTML]{FFCC67}\textbf{Medium} &
  \begin{tabular}[c]{@{}l@{}}Static: \\ Exhaustive analysis \\ and sanitisation of all \\ unvetted training \\ dataset and data source\end{tabular} \\ \hline
\end{tabular}%
}
\caption{The threat matrix for the University Virtual Assistant use case reflects the calculated likelihood, impact, and risk rating for prompt injection and training data poisoning following the completion of the risk analysis.}
\label{tab:threat-matrix-uni}
\end{table}

Enhancements in component security and robustness typically correlate with a decrease in the likelihood of successful attacks. However, it is important to maintain the independence of threat agent and impact factors from such changes, as adherence to worst-case scenarios mitigates the risk of oversight.  For instance, the implementation of a more robust prompt validation and filtering library, hardened with multi-layered prompt injection detection mechanisms, alongside response filtering to preempt potential prompt injection responses, can potentially increase the level of difficulty associated with prompt injection attacks. This would indeed decrease the likelihood of an attack occurring. However, if an adversary were to surpass these constraints, the impact would remain unchanged. Hence, in evaluating the impact, we solely consider the potential outcomes if the threat agent were to successfully exploit the vulnerability.

\section{Related work}

There are two primary avenues of research within the field of LLM security. The majority of studies introduce innovative techniques for compromising language models, while counter-studies aim to effectively mitigate or safeguard against such attacks. However, there is a scarcity of research focusing on the risk assessment procedures necessary for developers or security practitioners to evaluate the security resilience of their systems. Notable works falling into the former category include those from MITRE and OWASP.

MITRE's Adversarial Threat Landscape for Artificial-Intelligence Systems (ATLAS) serves as a structured knowledge base detailing adversary tactics, techniques, and procedures (TTPs) tailored specifically to AI systems. ATLAS offers valuable insights into potential AI security threats by delineating how adversaries might exploit vulnerabilities throughout the AI development and deployment lifecycle. These TTPs are elucidated through real-world case studies and attack illustrations, enhancing both comprehension and practical applicability. Additionally, brief mitigation strategies are provided to aid in addressing identified threats.

Cui et al.\cite{cui2024risk} classified risks associated with an LLM system into four modules: input, language model, toolchain, and output. They are also conducting risk assessments by evaluating the LLMs using benchmarking datasets to measure their robustness, truthfulness, and potential biases. Mauri and Damiani introduced STRIDE-AI \cite{mauri2022modeling}, a threat modeling approach based on the widely recognized STRIDE threat modeling from Microsoft. They utilized the Failure Modes and Effects Analysis (FMEA) process to identify potential failure modes within the ML lifecycle and map their causes and effects. Similar to our work, Wilhjelm and Younis explored the application of traditional threat modeling techniques for Machine Learning Based Systems (MLBS) \cite{wilhjelm2020threat}. They employed Data Flow Diagrams (DFDs) and STRIDE to identify threats, utilizing the Microsoft SDL AI/ML Bug Bar to rank threats based on their impact. Finally, they referenced the Microsoft AML attack library to propose mitigation strategies.

Q. Zou et al. proposed ML System Security Analysis (ML-SSA), a system architecture-centered security analysis comprising two main graphs \cite{zou2022attacks}. The first graph captures all cause-and-effect relationships relevant to assessing the likelihood of adversarial consequences in an ML system. The second graph delineates typical dependencies found in software systems, including those introduced by the supply chain perspective of an ML system. Berryville Institute of Machine Learning identified 78 generic ML risks and subsequently highlighted the top 10 risks from this set \cite{mcgraw2020architectural}. They categorised the components in a generic ML system and mapped them to associated risks.

Kapoor and Bommasani et al. \cite{openFM} anaylses the benefits and risks posed by open foundation model. They introduce a framework that focuses on the concept of \textit{marginal} risk, aiding in the identification of additional risks society faces due to open foundation models compared to pre-existing technologies or other relevant benchmarks. This framework involves several key steps: identifying the threat, comprehending the existing risk and its defenses in the absence of open foundation models, presenting evidence of the marginal risks posed by open foundation models, assessing the ease with which we can defend against these new risks, and finally outlining the  assumptions and uncertainties related to the risks under consideration. They use this framework to analyse past studies that analysed cyber risks possed by open foundation models and found them to be incomplete. Based on their study, they put forward recommendations to AI developers, researchers investigating AI risks, and policymakers and regulators.

\section{Conclusion and Future Work}

This study demonstrates the applicability of established risk assessment methodologies, such as the OWASP risk rating method, to the unique context of LLM-based systems. By combining scenario analysis, dependency mapping, and impact analysis, we provide a systematic framework for identifying and prioritising risks arising from LLM integration. The hypothetical use case showcases the practical utility of this approach, empowering security professionals and developers to make informed decisions about risk mitigation and resource allocation. In the scenario we assessed, we found that prompt injection presents a significant threat with a high-risk rating, whereas training data poisoning has a medium-risk rating. A security practitioner could utilize this information to prioritize mitigation efforts for prompt injection over training data poisoning. This might involve recommending adjustments to the prompt validation and filtering library being utilized, or implementing a more stringent LLM response filtering mechanism.

Our risk assessment process, culminating in the creation of the threat matrix, equips each stakeholder group with the knowledge to navigate LLM-related risks effectively.  This helps organisations understand the unique risks faced by each stakeholder group, enabling them to prioritize mitigation strategies and tailor security measures accordingly.

Despite the promise of LLMs, their associated risks and still-evolving nature highlight the need for a comprehensive security approach. We believe this work can serve as a foundation for more robust risk assessment practices within the rapidly developing field of LLM applications. Continuous refinement of the threat matrix will be necessary in response to emerging risks and changes in defensive and attack strategies. Real-world case studies evaluating the effectiveness of the proposed process and identifying potential shortcomings could further refine the process and improve its applicability.

In essence, our study serves as a foundation for ongoing discussions and advancements in securing LLMs, aiming to pave the way for a more resilient and trustworthy integration of these powerful language models in future applications.

\clearpage

\printbibliography

\end{document}